\documentstyle[preprint,aps]{revtex}
\def\mu{\lambda}
\def\Phi{U}

\begin{document}

\draft

% force linebreaks with \\
\title{
Ordering and Demixing  Transitions in Multicomponent Widom--Rowlinson Models
}

\author{
J.L. Lebowitz
}

\address{
Dept. of Mathematics and Physics, 
Rutgers University, New Brunswick, N.J. 08903, USA
}

\author{
A. Mazel
}

\address{
Dept. of Mathematics,  
Rutgers University, New Brunswick, N.J. 08903, USA
}
\address{
International Institute of Earthquake Prediction Theory and Mathematical
Geophysics, Moscow 113556, Russia
}

\author{
P. Nielaba 
}

\address{
Institut f\"ur Physik, Johannes--Gutenberg--Universit\"at,  
D--55099 Mainz, Germany
}

\author{
L. \v Samaj$^*$
}

\address{
Courant Institute, New York University, 251 Mercer Street, New York, 
N.Y. 10012--1185, USA
}

\date{\today}
\maketitle
\begin{abstract}
We use Monte Carlo techniques and analytical methods to study the phase
diagram of multicomponent Widom-Rowlinson models on a square lattice:
there are $M$ species all with the same fugacity $z$ and a nearest
neighbor hard core exclusion between unlike particles. Simulations show
that for $M$ between two and six there is a direct transition from the gas
phase at $z < z_d (M)$ to a demixed phase consisting mostly of one species
at $z > z_d (M)$ while for $M \geq 7$ there is an intermediate ``crystal
phase'' for $z$ lying between $z_c(M)$ and $z_d(M)$.  In this phase, which
is driven by entropy, particles, independent of species, preferentially
occupy one of the sublattices, i.e.\ spatial symmetry but not particle
symmetry is broken.  The transition at $z_d(M)$ appears to be first order
for $M \geq 5$ putting it in the Potts model universality class.  For
large $M$ the transition between the crystalline and demixed phase at
$z_d(M)$ can be proven to be first order with $z_d(M) \sim M-2 + 1/M +
...$, while $z_c(M)$ is argued to behave as $\mu_{cr}/M$, with $\mu_{cr}$
the value of the fugacity at which the one component hard square lattice
gas has a transition, and to be always of the Ising type.  Explicit
calculations for the Bethe lattice with the coordination number $q=4$ give
results similar to those for the square lattice except that the transition
at $z_d(M)$ becomes first order at $M>2$. This happens for all $q$,
consistent with the model being in the Potts universality class.
\end{abstract}

\pacs{PACS numbers: 64.60.Cn, 05.50.+q, 02.70.Lq, 75.10.Hk}

%\narrowtext
%\twocolumn
%\begin{multicols}{1}

%%%%%%%%%%%%%%%%%%%%%%%%%%%%%%%%%%%%%%%%%%%%%%%%%%%%%%%%%%%%%%%%%%%%%%%%%%%
%%%%%%%                       BODY OF TEXT
%%%%%%%%%%%%%%%%%%%%%%%%%%%%%%%%%%%%%%%%%%%%%%%%%%%%%%%%%%%%%%%%%%%%%%%%%%%

\section{Introduction}

In 1970 Widom and Rowlinson (WR) introduced an ingeniously simple model
for the study of phase transitions in continuum fluids~\cite{r1}.  It
consists of two species of particles, A and B, in which the only
interaction is a hard core exclusion between particles of unlike species,
i.e.\ the pair potential $v_{\alpha \beta}(r)$ is infinite if $\alpha \ne
\beta$, and $r < R_{AB}$, and is zero otherwise.  WR showed how this model
can be transformed (by integrating over the coordinates of one species)
into a one component model with explicit many body interactions.  The A-B
symmetry of the demixing phase transition in the original model, assumed
by WR to occur in dimensions $\nu \geq 2$ when the fugacity, $z_A = z_B =
z$ is large, then yields interesting information about the corresponding
liquid-vapor transition in the transformed model~\cite{r1}.

A rigorous proof of the existence of a demixing transition in this model
was given by Ruelle~\cite{r2}.  Ruelle used a brilliant adaptation of the
Peierls argument for the Ising model on a lattice which exploits the A-B
symmetry.  Further results were obtained in~\cite{r3}.  Ruelle's proof,
which permits also a smaller hard core $R_{AA} = R_{BB} < (\sqrt
{3}/2)R_{AB}$ between like particles, was generalized by Lebowitz and
Lieb~\cite{r4} to the case where $v_{AB}(r)$ is large positive but not
infinite.  An extension of the proof to non-symmetric multicomponent
models was made by Bricmont, Kuroda and Lebowitz using Pirogov-Sinai
theory~\cite{r4a}.  We refer the reader to~\cite{r4a} and references there
for additional results on these WR models which are, as far as we know,
the only continuum systems with fixed decaying potentials where one has
been able to prove rigorously the existence of phase transitions.

The lattice version of the multicomponent WR model---hard core exclusion
between particles of $M$ different species on nearest neighbor sites of a
simple cubic lattice in $\nu$-dimensions---is much easier to handle
rigorously.  A proof of the demixing transition and much more, e.g.\
existence of sharp interfaces between coexisting phases, in $\nu \geq 3$,
at large fugacity $z>z_d(M)$, can be obtained using standard Peierls
methods, see~\cite{r4b}, ~\cite{r6}.  A rather surprising result (at least
on first sight) was found by Runnels and Lebowitz~\cite{r9}.  They proved
that when the number of components $M$ is larger than some minimum $M_0$
then the transition from the gas phase at small values of $z$ to the
demixed phase at large values of $z$ does not take place directly.
Instead there is, at intermediate values of $z$, $z_c < z < z_d$, an
ordered phase in which one of the sublattices (even or odd) is
preferentially occupied, i.e.\ there is a crystalline
(antiferromagnetically ordered) phase in which the average particle
density on the even and odd sublattices, $\rho_{e}$ and $\rho_{o}$ are
unequal. The average density of species $I = 1,\dots,M,\; \rho(I)$, is the
same on each sublattice and equal $\rho_e(I) = M^{-1} \rho_e$ and
$\rho_o(I)= M^{-1} \rho_o$. The nature of the symmetry breaking is thus
very different from that in the demixed phase at $z > z_d$ where $\rho_{e}
= \rho_{o} = \rho$ but there exists one species, say $I'$, for which
$\rho(I') > M^{-1}\rho$.  The origin of the spatial symmetry breaking
leading to the crystal phase is purely entropic.  For $z$ fixed, and $M$
going to infinity it pays for the system entropywise to occupy just
one sublattice without any constraint; there being no interaction between
particles on the same sublattice each site can be occupied independently
by a particle of any species, i.e.\ if we keep one of the sublattices
empty then there are $M$ independent choices at each site of the other
sublattice.  This more than compensates, at $M > M_0$, for the ``loss'' of
fugacity occasioned by keeping down the density in the other
sublattice. Put in another way, for $M$ large enough, the typical
occupancy pattern on a lattice (ignoring the label $I$ of the particles)
should behave like a one component lattice gas with nearest neighor hard
core exclusion for which Dobrushin~\cite{r5} proved the existence of a
crystalline state.

Once this is understood the natural question is, just how big does $M_0$
have to be to see this ordered phase for $M \geq M_0$.  It was shown 
in~\cite{r9} that $M_0 < 27^6$; a ridiculously large upper bound.  On the other
hand application of Pirogov-Sinai theory as in ~\cite{CKS} where
a similar model, in which there is a positive energy 
$\Phi < \infty$ 
when neighboring sites are occupied by different species, is considered can
probably be made to give $M_0 \stackrel{<}{\sim} 25$ for our model
corresponding to 
$\Phi = \infty$. 
Furthermore, a direct computation on the Bethe lattice with $q$-neighbors,
gives $M_0 = [q/(q-2)]^2$, which would suggest $M_0
\sim 4$ for the square lattice.
The Monte Carlo (MC) simulations 
presented here grew out from a desire to answer this question and to get a 
general picture of the phase diagram of this system in the $M-z$ plane. 
This we succeeded in doing, with the large $M$ analytic results smoothly 
matching up with the MC small $M$ results and with the ``dilute 
lattice model'' investigated in~\cite{CKS}. The outline of the rest of the
paper is as follows. In Section II we investigate the large $M$ behavior
of the model, Bethe lattice computations are given in Section III and MC
simulations in Section~IV.

\section{asymptotics of the phase diagram}

We consider a two dimensional square lattice ${\bf Z}^2$. Each lattice
site can be either empty $(I=0)$ or singly occupied by a particle of type
$I=1,2,\cdots,M$.  All the components have the same activity $z$ and there
is an infinite repulsive interaction between particles of different type
on nearest neighbor sites.  Thus a particle of type $I$ at lattice site
$i$ can only have vacancies or particles of type $I$ on nearest neighbor
lattice sites. The interaction potential $\phi_{I,J}(i,j)$ between a
particle of type $I$ at site $i$ and a particle of type $J$ at site $j$
is:

\begin{equation}
\phi_{I,J}(i,j) = \left\{
\begin{array}{l}
\infty\ {\rm if}\ i\ {\rm and}\ j\ {\rm are\ nearest\ neighbors\ and\ }
I\neq J, I\neq 0, J\neq 0\;
\\
0 \ {\rm otherwise}
\end{array}\right. \label{eq1}
\end{equation}

It is clear that if we replace $\infty$ in (\ref{eq1}) by some 
$\Phi \neq 0$ then our system is equivalent to a dilute Potts model. In
such a system
some lattice sites are empty while others are occupied, with a weight given
by the fugacity $z$, by an $M$-component Potts variable with nearest
neighbor interaction between like states equal to
$-\Phi$.
The WR system (\ref{eq1}) can thus be
considered as the zero temperature limit of such a model with 
$\Phi > 0$. We refer the reader to~\cite{CKS} for a general discussion of
the phase diagram of such models.  
%%%%%%%%%%%%%%%%%%%%%%%%%%%%%%%%%%%%%%%%%%%%%%%%%%%%%%%%%%%%%%%%%%%%%%%%%%%
%%%%%%%                       Text by Mazel
%%%%%%%%%%%%%%%%%%%%%%%%%%%%%%%%%%%%%%%%%%%%%%%%%%%%%%%%%%%%%%%%%%%%%%%%%%%

\def\s{\sigma}
\def\d{\delta}
\def\z#1{{\bf Z}^{#1}}

\subsection{The Boundary Between Disordered and Crystal Phases}

We will argue here that the asymptotics, as $M \to \infty$, of
the boundary $z=z_c(M)$ between disordered and crystal
phases is given by the hyperbola $Mz=\mu_{cr}$, where $\mu_{cr}$ is the
critical fugacity of the one-component lattice gas with the
nearest-neighbor hard-core exclusion.

In the latter model every site of the lattice can be occupied by a
particle with fugacity $\mu>0$. Hard-core repulsion requires that
occupied sites are not nearest neighbors. It is well known~\cite{BET} that
for this model there exists a critical fugacity $\mu_{cr}\sim 3.7962\;$
such that for $\mu < \mu_{cr}$ the model posseses a unique limit Gibbs
state (disordered phase) while for $\mu > \mu_{cr}$ it has at least two
different limit Gibbs state (crystal phases).  In one of the crystal
phases the probability of the even site to be occupied by a particle is
greater than that of the odd site and vice versa for the other crystal
phase.

Consider now a model with $M$ different species in which we forbid
particles of any type to be nearest neighbors.  Then, for fixed $\mu=Mz$,
this system is equivalent to the one-component lattice gas just described.
Hence the multi-component WR system, where particles of the same type can
occupy neighboring lattice sites is, in a sense, the one-component system
with the hard-core condition being slightly relaxed. That naturally leads
to a conjecture that for $Mz\le \mu_{cr}$ our multicomponent system is in
the disordered phase.

Speaking more precisely, any configuration of the multicomponent system
(\ref{eq1}) can be uniquely decomposed on the connected components of the
occupied sites.  The connected component consisting of a single site can
be interpreted as a hard-core particle with the fugacity $Mz$ as we do not
specify the type of the particle. Other connected components consisting of
$n \ge 2$ sites we call clusters. All the particles in the cluster are of
the same type and the fugacity of the cluster is $Mz^n$, as we again do not
specify the type of the particles in the cluster. Note that this
representation naturally extends the multicomponent model to non integer
values of $M$.

It is obvious that the fugacity of clusters tends to 0 as $z \to 0$,
$\mu=Mz$ fixed, and the free energy of the multicomponent model tends to
the free energy of the one-component hard-core gas with the fugacity
$\mu$, i.e. the contribution of the clusters becomes negligible in the
limit $z \to 0$, with $\mu=Mz$ fixed. This suggests that in the $(M,z)$ plane
there exists a curve $M=M_c(z)$, with $M_c(z) \to \mu_{cr} / z$ as $z \to
0$, on which there is the second order phase transition between disordered
and crystal phases. The typical configuration of the crystal phase has one
sublattice, say the even one, occupied by particles except for
rare excitation "islands", where every "island" is either a cluster or has
the structure of the opposite crystal phase with occupied odd sublattice.

\subsection{The Boundary Between Crystal and Demixed Phases}

The boundary between crystal and demixed phases admits a rigorous analysis
in the framework of the Pirogov-Sinai theory (see \cite{PS}, \cite{S},
\cite{Z}). Related general results for the wider class of models
including our multi-component system can be found in \cite{CKS}. Here we
present a more detailed analysis by means of a direct "elementary" approach.

Given a configuration of the multi-component system we say that this
configuration at a site of the lattice $\z{2}$ is,

{(i)} in the demixed phase $I$, if this site and at least one of its
nearest neighbor sites are occupied by particles of type $I$;

{(ii)} in the odd crystal phase, if this site is even and empty or is
odd and occupied by a particle (of any type) while all four nearest
neighbor sites are empty;

{(iii)} in the even crystal phase, if this site is odd and empty or  is
even and occupied by a particle (of any type) while all four nearest
neighbor sites are empty.

\medskip\noindent
It is not hard to check that the statistical weight of an arbitrary
configuration can be calculated as the product over all unit bonds of the
lattice $\z{2}$ of the following statistical weights of the bonds:

{(i)} $z^{1/2}$ for the bond joining two sites in the same demixed
phase;

{(ii)} $(Mz)^{1/4}$ for the bond joining two sites in the same
crystal phase (necessarily one of the sites is occupied and other is
empty);

{(iii)} $1$ for the bond joining sites in the different crystal phases
(necessarily both sites are empty);

{(iv)} $z^{1/4}$ for the bond joining sites in the crystal and
demixed phases (the site in the crystal phase is necessarily empty).

\medskip\noindent All other bonds are forbidden, i.e. they have a zero
statistical weight.  Without changing the Gibbs distribution we multiply
all statistical weights of the bonds by $z^{-1/2}$ and obtain renormalized
statistical weights $1$, $(M/z)^{1/4}$, $z^{-1/2}$ and $z^{-1/4}$
respectively. From that picture it is clear that for $1 \le M <z$ the
configurations with all sites being in the same demixed phase are the only
periodic ground states (i.e. the configurations minimizing the specific
energy) of the model (there are $M$ of them). For $0< z < M$ the
configurations with all sites being in the same crystal phase are the only
periodic ground states of the model (there are $2$ of them). Finally on
the line $M=z$ all $M+2$ configurations above are the ground states of the
model and there is no other periodic ground state.

Given a configuration we introduce the Peierls contours as the connected
components of the unit bonds of the dual lattice ${\widetilde{\bf Z}}^2
%\z{2}
=\z{2}+(1/2,1/2)$ separating sites of $\z{2}$ not in the same
phase. It is not hard to see that every configuration has an equivalent
representation in terms of a collection of mutually disjoint Peierls
contours.  Moreover, on the line $M=z$ the statistical weight of the
configuration is the product of the statistical weights of the
contours. In turn, the statistical weight of a contour is the product of
the statisical weights of the bonds of $\z{2}$ dual to the bonds of the
contour. Here dual means rotated by $\pi / 2$ with respect to the
center of the bond. This representation is customary for the Pirogov-Sinai
theory and allows complete investigation of the diagram of the periodic
limit Gibbs states in the region $M \ge \max (z_0/z, 1)$ for a sufficiently
large absolute constant $z_0$.

{\bf Theorem.} {\sl For $M \ge \max (z_0/z, 1)$ and $z_0$ large enough
there exists in the $(M,z)$ plane a curve $M=M_d(z),\ |M_d(z)-z|=O(1)$,
such that on this curve all $M+2$ ground states generate the corresponding
limit Gibbs measures and these limit Gibbs measures are the only periodic
limit Gibbs measures.

For $\max (z_0/z, 1) <M<M_d(z)$ every demixed ground state generates the
corresponding limit Gibbs state and these $M$ limit Gibbs states are the
only periodic limit Gibbs measures.

For $M>\max (z_0/z, 1, M_d(z))$ every crystal ground state generates the
corresponding limit Gibbs state and these $2$ limit Gibbs states are the
only periodic limit Gibbs measures.}

\medskip\noindent
(See \cite{PS}, \cite{S}, \cite{Z} and cf. \cite{CKS}).

More precise asymptotics $M_d(z)=z+2-z^{-1}+O(z^{-2})$ as $z \to \infty$
can be calculated via an approach suggested in \cite{Sl}. Namely, set
$M_d(z)=z+\sum_{n=0}^{\infty} c_n z^{-n}$. Then one can successively
calculate $c_n$ by equating term by term the cluster or polymer expansion
series written for the specific free energies $f_c$ and $f_d$ of the
crystal and demixed phases respectively. The definition of polymers and
their statistical weights can be found in \cite{Se}.

It is not hard to see that 
$$f_d=-\log z -z^{-1}+{1 \over 2} z^{-2}+O(z^{-3}),$$ 
where

{(i)} $-\log z$ is the specific energy of the demixed ground state;

{(ii)} $z^{-1}$ is the statistical weight of the polymer consisting of
a single excitation obtained from the demixed ground state by
removing a particle from a given lattice site;

{(iii)} $z^{-2}/2$ is the statistical weight of the polymer
consisting of two copies of the excitation defined in (ii);

{(iv)} $O(z^{-3})$ is the contribution of the rest of polymers.

\medskip\noindent
Similarly 
$$
f_c=-{1 \over 2} \log (Mz) -{1 \over 2} (Mz)^{-1}-{1 \over2}M (zM^{-4}) +
O(z^{-3})$$
$$= -\log z - {c_0 \over 2} z^{-1} - \left({c_1 \over2}- {c_0^2
\over 4} +1 \right) z^{-2} + O(z^{-3}),$$ 
where

{(i)} $-[\log (Mz)]/2$ is the specific energy of the crystal ground state;

{(ii)} $(Mz)^{-1}$ is the statistical weight of the polymer consisting of
a single excitation obtained from the crystal ground state by
removing a particle from an occupied site; the factor $1/2$ is due to the
fact that in the crystal ground state only one half of the lattice points
are occupied;

{(iii)} $zM^{-4}$ is the statistical weight of the polymer consisting of
a single excitation obtained from the crystal ground state by
placing a particle of a given type $I$ at a previously non occupied
lattice site; such a particle requires the neighboring sites to be in the
same phase $I$; any of $M$ types of particles can produce such an
excitation which is reflected in the factor $M$; finally, the factor $1/2$
is due to the fact that in the crystal ground state only one half of the
lattice points is non occupied;

{(iv)} $O(z^{-3})$ is the contribution of the rest of polymers;

\medskip\noindent 
Thus equating $f_d$ and $f_c$ one obtains $c_0=2$ and $c_1=-1$.  Note that
every $c_n$ can be calculated in the same way but the amount of
calculation grows very fast with $n$.
%%%%%%%%%%%%%%%%%%%%%%%%%%%%%%%%%%%%%%%%%%%%%%%%%%%%%%%%%%%%%%%%%%%%%%%%%%%
%%%%%%%              End of Text by Mazel
%%%%%%%%%%%%%%%%%%%%%%%%%%%%%%%%%%%%%%%%%%%%%%%%%%%%%%%%%%%%%%%%%%%%%%%%%%%

\section{The Bethe lattice computation}

\def\Omega{\Xi}

We now present an exact calculation of the phase diagram for our
multicomponent WR model formulated on the Bethe lattice of general
coordination number $q$.  The computation is based on the exact ``inverse
solution'' for simply connected lattice
structures~\cite{r12},~\cite{r12a}.  The approach is a more transparent
alternative to the usual recursion method, it replaces the most stable
fixed point criterion by the general principle of the global minimum of
the free energy.

In the Bethe lattice, every vertex $i$ is an articulation
point of multiplicity $q$.  Let us denote by $\{ z_i(I), I=0,1,\ldots,M
\}$ with the reference $z_i(0) \equiv 1$ the set of fugacities assigned to
the corresponding particle states at site $i$, and by $\{ \rho_i(I) \}$
the generated particle-density field constrained by 
\begin{equation}
\sum_{I=0}^M \rho_i(I) = 1.
\label{LSeq0}
\end{equation}
A direct method assumes given fugacities $\{ z_i(I) \}$ and then
calculates the densities $\{ \rho_i(I) \}$ and the specific free energy via
the Gibbsian grand canonical ensemble formalism. In the inverse method on
the contrary we calculate the specific free energy and fugacities $\{
z_i(I) \}$ as functions of densities $\{ \rho_i(I) \}$ considered as the
basic variables. The articulation character of vertices in the Bethe 
lattice then permits the topological reduction of the equilibrium
description. In particular:

(i) the inverse profile equation, i.e. the dependence of the fugacity
field $\{ z_i(I) \}$ on the specified density field $\{ \rho_i(I) \}$,
takes the local form~\cite{r12}
\begin{equation}
z_i(I) = \left[ {\rho_i(0)\over \rho_i(I)} \right]^{q-1}
\prod_{j=1}^q z_i^{<i,j>}(I) \;.
\label{LSeq1}
\end{equation}
Here and later, the superscript $<i,j>$ refers to the model (1) on an
isolated (i.e. with empty boundary conditions) bond joining two
neighboring sites $i$
and $j$. The partition function of this two-site model is denoted by
$\Omega^{<i,j>}$. The fugacities are denoted by $z_i^{<i,j>}(I)$ and
$z_j^{<i,j>}(I)$. The densities are $\{ \rho_i(I) \}$ and $\{ \rho_j(I)
\}$. In the notation for densities the superscript is omitted as these
densities coincide with actual ones on the Bethe lattice which  we always
assume to be {\it specified}.

(ii) The Helmholtz free energy of a system expressed via densities, $F[\rho]= - \log \Omega
+\sum_{i,I}\rho_i(I) \log z_i(I)$ with $\Omega$ being the grand partition
function, can be calculated as (see~\cite{r12a})
\begin{equation}
F[\rho] = \sum_{<i,j>} F^{<i,j>}[\rho_i,\rho_j] - (q-1)\sum_i
F^i[\rho_i] \;,  
\label{LSeq3}
\end{equation}
where the summation over $<i,j>$ is over every pair of nearest
neighbors. The one-site and the two-site Helmholtz free energies are
\begin{equation}
  F^i[\rho_i] =-\log \Omega^i +\sum_{I=0}^M \rho_i(I)\log z_i(I)=
\sum_{I=0}^M \rho_i(I) \log \rho_i(I) \;  
\label{LSeq4}
\end{equation}
and 
\begin{equation}
  F^{<i,j>}[\rho_i,\rho_j] = -\log \Omega^{<i,j>}+\sum_{I=0}^M
\rho_i(I)\log z_i(I) +\sum_{I=0}^M \rho_j(I)\log z_j(I).
\label{LSeq4a}
\end{equation}

In order to mimic realistically a non-simply connected structure of the same
coordination $q$, it is necessary to avoid the effect of the large number of
boundary sites of the Cayley tree.  This can be done by assuming that, in
the thermodynamic limit, the local properties of interior vertices
(expressed in terms of particle densities) are equivalent, i.e. by
considering only translation-periodic extremal Gibbs measures on the
Cayley tree. Under homogeneous external conditions $z_i(I) =
z(I)$ for all $i$, the existence of a phase transition can be simply
detected from~(\ref{LSeq1}) as a bifurcation point, resp. a ``jump'' point
for the densities, with the minimum principle of the free energy,
determining the right equilibrium.
\subsection{Crystal Phase}

The crystal regime of our model is characterized by the supposed
symmetry-breaking in particle densities on alternating
sublattices. Accordingly we set $\rho_i(I)=\rho_1$ for sites $i$
on the first sublattice and $\rho_j(I)=\rho_2$ for sites $j$ on
the second sublattice $(I=1,\ldots,M)$. In the corresponding two-site
model we suppose that $z_i^{<i,j>}(0)=z_j^{<i,j>}(0)=1$, $z_i^{<i,j>}(I) =
z_1$ and $z_j^{<i,j>}(I) = z_2$ 
$(I=1,2,\ldots,M)$ since there is no preference for any component at a
given sublattice.  With this notation the two-site pair partition function $
\Omega^{<i,j>}$ is

\begin{equation}
  \Omega^{<i,j>} = 1 + M (  z_1 +   z_2) + M   z_1   z_2 \; .
\label{LSeq5}
\end{equation}
As functions of $  z_1$ and $  z_2$ the one-site particle
densities $\rho_1$ and $\rho_2$ 
are given by
\begin{mathletters}
\label{LSeq6}
\begin{equation}
\rho_{1}   \Omega^{<i,j>} =   z_{1} (1 +   z_{2}) \; .
\end{equation}
\begin{equation}
\rho_{2}   \Omega^{<i,j>} =   z_{2} (1 +   z_{1}) \; .
\end{equation}
\end{mathletters}
>From (\ref{LSeq5}), (\ref{LSeq6}) we easily get $  \Omega^{<i,j>}$ and $ 
z_{1,2}$ as functions of  $\{ \rho_1, \rho_2 \}$:
\begin{mathletters}
\label{LSeq7}
\begin{equation}
  z_{1} = \frac{M(\rho_{1}+\rho_{2}) -1 + \rho_{1} - \rho_{2}
+ D^{1/2}}{2(1-M \rho_{1})} \;, 
\end{equation}
\begin{equation}
  z_{2} = \frac{M(\rho_{2}+\rho_{1}) -1 + \rho_{2} - \rho_{1}
+ D^{1/2}}{2(1-M \rho_{2})} \;, 
\end{equation}
\end{mathletters}

\begin{equation}
  \Omega^{<i,j>} = \frac{M(M-1) (\rho_1+\rho_2) - (M-2) + M D^{1/2}}
{2(1-M \rho_1) (1-M \rho_2)} \;, 
\label{LSeq8}
\end{equation}
where the plus sign of the square root of the discriminant
\begin{equation}
D = \left[ 1 - (M-1) (\rho_1+\rho_2) \right]^2 + 
4 \rho_1 \rho_2 (M-1) 
\label{LSeq9}
\end{equation}
is fixed by the condition $ z_{1,2}\to 0$ for $\rho_{1,2}\to 0$.  Finally,
using (\ref{LSeq0}) and (\ref{LSeq1}) with homogeneous external conditions
$z_i(I) = z$ for all $i$ and $I=1,\ldots,M$ , we find
\begin{mathletters}
\label{LSeq10}
\begin{equation}
z = \left( {1-M \rho_{1} \over \rho_{1}} \right)^{q-1} 
  z_{1}^q \;,
\end{equation}
\begin{equation}
z = \left( {1-M \rho_{2} \over \rho_{2}} \right)^{q-1} 
  z_{2}^q \;.
\end{equation}
\end{mathletters}
Because of translation periodicity the free energy per site, $f_c$, can be
determined from (\ref{LSeq3}) as follows
\begin{equation}
f_c = - {q\over 2} \log   \Omega^{<i,j>}
-  {q-1\over 2}  \log \left[ (1-M \rho_1) (1-M \rho_2) \right]. 
\label{LSeq11}
\end{equation}

The relations (\ref{LSeq10}) can be considered as the equation
$z(\rho_1,\rho_2)=z(\rho_2,\rho_1)$ (~$0~< \rho_1,\rho_2 < M^{-1}~)$. In
the variables $s=(\rho_1+\rho_2)/2$ and $t=\rho_1-\rho_2$ it can be rewritten 
as $z(s,t)=z(s,-t)$ and always has a trivial solution (s,0). There is
no other solution if $s \in [0,s^L_c) \cup (s^U_c, M^{-1}]$,
\begin{mathletters}
\label{LSeq12}
\begin{equation}
s_c^L = {1\over 2M} (1 - E^{1/2}) \;, \label{LSeq12a}
\end{equation}
\begin{equation}
s_c^U = {1\over 2M} (1 + E^{1/2}) \;, \label{LSeq12b}
\end{equation}
\end{mathletters}
and
\begin{equation}
E = 1 - {4 M (q-1) \over (M-1) q^2} \;, \label{LSeq13}
\end{equation}
because for these $s$, calculated from the equation ${\partial
z}(s,0)/{\partial t}=0$, the derivative ${\partial z}(s,t)/{\partial t}$
has a constant sign. For $s \in (s^L_c,s^U_c)$ one has a nontrivial
solution to $z(s,t)=z(s,-t)$. The corresponding critical fugacities read
\begin{mathletters}
\label{LSeq14}
\begin{equation}
z_c^L = M^{q-1} \left( {1-E^{1/2}\over 1+E^{1/2}} \right)
\left[ {(q-2)/q - E^{1/2}\over 1 - E^{1/2}} \right]^q \;, \label{LSeq14a}
\end{equation}
\begin{equation}
z_c^U = M^{q-1} \left( {1+E^{1/2}\over 1-E^{1/2}} \right)
\left[ {(q-2)/q + E^{1/2}\over 1 + E^{1/2}} \right]^q  \;. \label{LSeq14b}
\end{equation}
\end{mathletters}
The critical point exists provided that $E\ge 0$, which imposes
the requirement on the number of components $M \ge M_0$ with the
minimum value $M_0$ given by
\begin{equation}
M_0 = \left( {q\over q-2} \right)^2 \;. \label{LSeq15}
\end{equation}
It is easy to verify, using (\ref{LSeq11}), that the crystal phase is
thermodynamically dominant (i.e. it has the minimal specific free energy)
with respect to the disordered one in the interval of activities $z_c^L <
z < z_c^U$. It is not hard to see that for large $M$
\begin{equation}
z_c^L \sim \frac{1}{M}
\frac{(q-1)^{q-1}}{(q-2)^q}=\frac{\lambda^{Bethe}_{cr}}{M} 
\label{LSeq15a}
\end{equation}
and
\begin{equation}
z_c^U \sim M^{q-1} \frac{(q-2)^q}{(q-1)^{q-1}}.
\label{LSeq15b}
\end{equation}

\subsection{Demixed Phase}

In the demixed phase regime, the sites are equivalent but one of
components, say $I=1$, has greater density: $\{ \rho_i(1)=\rho(1),
\rho_i(I) = \rho(2)$ for all $I = 2,\ldots,M \}$. Assuming
$z_i^{<i,j>}(0)=z_j^{<i,j>}(0)=1$ and denoting
$z_i^{<i,j>}(1)=z_j^{<i,j>}(1)=z(1)$, $z_i^{<i,j>}(I)=z_j^{<i,j>}(I)=
z(2),\ I = 2,\ldots,M$ for the two-site model we have
\begin{equation}
  \Omega^{<i,j>} = [1+  z(1)]^2 + (M-1)[1+  z(2)]^2
- (M-1) \;, \label{LSeq16}
\end{equation}
\begin{mathletters}
\label{LSeq17}
\begin{equation}
\rho(1)   \Omega^{<i,j>} =   z(1) [1+  z(1)] \;.
\end{equation}
\begin{equation}
\rho(2)   \Omega^{<i,j>} =   z(2) [1+  z(2)] \;.
\end{equation}
\end{mathletters}
The system of equations (\ref{LSeq1}) is closed by considering
$z_i(I)=z,\; I=1,\ldots,M$ and~(\ref{LSeq0}), which leads to the equations
\begin{mathletters}
\label{LSeq18}
\begin{equation}
z = \left[ {1-\rho(1)-(M-1)\rho(2) \over \rho(1)} \right]^{q-1}
  z(1)^q \;.
\end{equation}
\begin{equation}
z = \left[ {1-\rho(1)-(M-1)\rho(2) \over \rho(2)} \right]^{q-1}
  z(2)^q \;.
\end{equation}
\end{mathletters}
The free energy per site is readily obtained in the form
\begin{equation}
f_d = -{q\over 2} \log   \Omega^{<i,j>} - (q-1) \log
[1-\rho(1)-(M-1)\rho(2)] \;.  
\label{LSeq19}
\end{equation}

We solved the non linear equations (\ref{LSeq16})-(\ref{LSeq18})
numerically. Given $M$, there is for small $z$ only one solution
corresponding to the disordered phase. As we increase $z$, two other
solutions appear at some $z=\bar z_d(M)$. For bigger values of $z$ even
more solutions exist. It appears that among these nontrivial solutions the
solution with maximal $\rho(1)$ always has the minimal free energy. We
call it demixed 1 (d1) solution. Finally we calculate the phase diagram
for $q=4$ (see Fig. 1) by comparing the free energies of disordered,
crystal and demixed 1 phases.

The behavior of the solutions is very similar to that of the zero-field
Potts model on the Bethe lattice~\cite{r12b,r12c} (where the role of
fugacity is played by the coupling constant). More precisely,
simultaneously with d1 another solution, d2, appears at the point $\bar
z_d(M)$ (see Fig.~2a). At $\bar z_d$, the free energy of d1 and d2 phases
is greater than that of the disordered phase and
$\rho_{d1}(1)=\rho_{d2}(1)>\rho_{dis}(1)$.  Increasing $z$ further, these
two phases split and $\rho_{d1}(1)>\rho_{d2}(1)$.  At $z=z_d$ (or
$z=z_c$), the free energy of the d1 phase coincides with the one of the
disordered (or crystal) phase, and the system exhibits a first-order phase
transition, accompanied by a jump in densities $\rho_{dis}(1) \rightarrow
\rho_{d1}(1)$.  To complete the description, we mention that at a larger
value of $z$, $z=\tilde z_d$, given by
\begin{equation}
\tilde z_d = \left( {M-2+q \over q-1} \right)^{q-1} {1\over q-2}\;, 
\label{LSeq20}
\end{equation}
we have $\rho_{d2}(1)=\rho_{d2}(2)=\rho_{dis}$ and the free energies of
the disordered and d2 phases coincide.  For $z>\tilde z_d$, one observes
that $\rho_{d2}(1) < \rho_{d2}(2)$, i.e. one particle components is
paradoxically suppressed by the others for d2 solution which has in this
region a lower free energy than that of disordered phase but greater than
that of d1 phase.  The only exception from the above scenario is
represented by the $M=2$ component WR model (Fig.~2b).  In that case, the
d1 and d2 phases are in fact the equivalent realizations of the particle 1
$\leftrightarrow$ 2 exchange symmetry of the same demixed phase, and the
corresponding demixing phase transition is of second order.

Let $M_c(\nu)$ denotes the ``critical'' number of components of the WR
model in $\nu$-dimensions, such that the phase transition from the
disordered to the demixed phase is second order for $M \le M_c$ and first
order for $M>M_c$.  For the Bethe lattice, we have $M_c = 2$ independently
of the coordination number.  This value of $M_c$ is exactly the same as
the value of its counterpart defined for the ordinary zero-field $M$-state
Potts model on the Bethe lattice.  The equality $M_c = 2$ for the Potts
model is supposed to hold for regular lattices in dimensions $\nu \ge 4$,
where the mean-field treatment provides an adequate description of the
critical behavior.  We therefore suggest that our $M$-component WR model
is a dilute version of the $M$-state Potts model, preserving the $Z_M$
symmetry among the particle states, which falls into the same universality
class.  This conjecture is supported by the MC estimate $M_c = 4$ for the
WR model on the $\nu=2$ square lattice in accordance with the behavior of
the Potts model on the square lattice~\cite{Baxter}. This is discussed in
the next part of the paper.

\section{Monte Carlo Simulation}

In this section we present results of the Monte Carlo study of the $M$--
component WR model on a square lattice of size $S^2 = 100 \times 100$
with periodic boundary conditions.  On an initially empty lattice we
deposit particles chosen at random from the $M$ components at fugacity $z$
respecting the exclusion given by~(\ref{eq1}).  We sequentially update the
lattice using a checkerboard algorithm resulting in a good
vectorization. An update of a lattice site $(i_1,i_2)$ occupied by a
particle of type $I$ ($I=0$ indicating an empty site)
is done as follows: We randomly choose a new trial particle of type $I_{tr}$, 
where $I_{tr}$ can have any 
integer value between $0$ and $M$ with equal probability. 
$I_{tr} = 0$ refers to a removal attempt 
of a particle $I \not= 0$ from the lattice site, 
which is successful, if a number $X$ randomly chosen with equal
probability between $0$ and $1$ is smaller than the inverse fugacity
$1/z$.
In this case $I$
gets the value $0$, otherwise it remains unchanged.  $I_{tr} \neq 0$
refers to a deposition attempt of a particle of type $I_{tr}$.  If $I=0$
then it is successful if each of the four nearest neighbor sites is either
empty or occupied by a particle of the same type ($I_{tr}$) and $X <
z$. In this case $I$ gets the value $I_{tr}$, otherwise it remains
unchanged. A direct replacement attempt of a particle $I\not=0$ surrounded
by four empty nearest neighbor sites is always successful.  Typically in a
simulation run after an equilibration of $5\times 10^5$ Monte Carlo steps
(MCS) we update the lattice $5\times 10^5$ times (in the cases $M=6$ and
$z=2.5$, $3.$, $3.5$ and $4$ up to $5\times 10^6$ times), the
configuration of every tenth step is taken for the evaluation of the
averages. A typical run with $5\times 10^5$ MCS took about 3~CPU hours on
a CRAY--YMP.

Let $m(i_1,i_2)$ denote the occupancy of a site, $m(i_1,i_2) = 0$ if
the site $(i_1,i_2)$ is empty and
$m(i_1,i_2)=1$ otherwise.
As observables we took histograms $P_L(\phi_c)$ 
of the order parameter $\phi_c$ for the crystal structure
and $P_L(\phi_d)$ of the order parameter $\phi_d$ for
the demixed phase in subsystems of size $L\times L$,
\begin{equation}
\phi_c = \frac{1}{L^2} \sum_{i_1, i_2 = 1}^L 
\left[2 m(i_1,i_2)-1 \right](-1)^{i_1 + i_2}
\label{pc}
\end{equation}
and
\begin{equation}
\phi_d = \frac{1}{L^2} {\rm Max}_I N_L(I) -\rho/M
\label{pd}
\end{equation}
where 
$N_L (I)$ denotes the number of particles of type $I$ in a subsystem of
size $L \times L$ and $\rho$ is the average overall density.

In Fig.~3 we show typical configurations for $M=9$ at three 
different values of $z$.
In Fig.~3(a),  $z=0.1$, in this case the system is in the 
gas (or disordered phase). In Fig~1(b),  $z=5$, and the system is in the
crystal phase where one of the sublattices has a  higher density.
In Fig.~3(c), $z=8.5$ and the  lattice is
predominantly occupied by particles of one type (demixed regime).

We first discuss the phase transition from the gas to the 
crystal phase. 
For a given $M$ 
the transition activity $z_c$ is found
by finite size scaling techniques~\cite{PRIV,BIND1}. 
In particular the $k-th$ moments of the order parameter distribution
$P_L(\phi_c)$,
\begin{equation}
< \phi_c^k >_L:= \int \phi_c^k P_L (\phi_c) d\phi_c
\label{cmoments}
\end{equation}
can be evaluated in subsystems of size $L \times L$, and from them the
fourth order cumulant~\cite{BIND1} $U_L$\ \ ,

\begin{equation}
U_L = 1 - \frac{<\phi_c^4>_L}{3< \phi_c^2 >_L^2}\ .
\label{ccumu}
\end{equation}
In a one phase region far away from a critical point the subsystem size
typically can be chosen larger than the correlation length $\xi$, $L >>
\xi$ and the order parameter distribution is to a good approximation a
Gaussian centered around $0$, resulting in $U_L \to 0$ for $L \to \infty$.
In the two phase coexistence region far away from a critical point we can
again assume $L >> \xi$ and the order parameter distribution is bimodal
resulting in $U_L \to 2/3$ for $L \to \infty$.  Near the critical point
however we have $ L << \xi$, and using scaling arguments~\cite{BIND1} the
cumulant is a function of $L/ \xi$, resulting for $\xi \to \infty$ in the
same value of $U_\ast$ for all different $L$.  This method efficiently
allows the localization of critical points by analyzing the cumulants for
different values of $z$ on different length scales $L$. For low values of
$z$ we are in the disordered one phase region, here $U_{L'} > U_L$ for $L'
< L$.  For large enough $M$ we obtain a crystal phase with $U_{L'} < U_L$
for $L' < L$.  Near a critical point we should expect $U_{L'} \approx U_L$
for $L' \neq L$.

In Figs.~4-6 we present the results for $U_L$ as a function of
$z$. For $M=9$ we observe in Fig.~4 a cumulant intersection 
near $z_c=0.85 \pm 0.05$
indicating the phase transition from the disordered to the crystal phase.
For $M = 8$ we have an intersection point at $z_c=1.1 \pm 0.05$ (see Fig.~5)
and for $M =7$ at $z_c=1.6 \pm 0.1$ (see Fig.~6). For $M=20, 15, 14$  and $10$ 
we obtain $z_c = 0.24, 0.35, 0.38, 0.68 \pm 0.01$ respectively.
The transition points are presented in Fig.~7 together with
the asymptotic expression~\cite{r11} $Mz = 3.7962$, (section~II)
and the results from the Bethe lattice computation, (section~III).
We find good agreement of our MC results with those 
of the asymptotic form for large values of $M$
and for large $z$ with the results of the Bethe lattice computation as well.

The case $M = 6 $ is analyzed in Fig.~8.  In all cases studied the order
parameter cumulant is decreasing with increasing system size, see
Fig.~8(a), indicating the presence of the disordered phase. In Fig.~8(b)
we show the cumulant for a given system size versus $z$, no cumulant
intersection points were found in these cases, but $U_{L'} > U_L$ for $L'
< L$ even for $z$-- values as large as $z = 4.4$ indicating that the
system with $M = 6$ components is in the disordered phase.  Our conclusion
of the data analysis is that, based upon the present statistical effort,
no sign of a phase transition from the gas phase to the crystal phase was
found for $M \leq 6$, indicating the nonexistence of this transition for
$M \leq 6$.  Thus the minimum number of components required for the
existence of the crystal phase is $M=7$; with possibly a noninteger value
$M_0$ between $6$ and $7$.

The transitions from the disordered to the crystal phase were analyzed
further by finite size scaling techniques.  We plotted the scaling
functions of the order parameter and the order parameter susceptibility,
$\tilde{\phi_c} = L^{\beta/\nu} <|\phi_c|>_L$ and $\tilde{\chi_c} =
L^{2-\gamma/\nu} \left[<\phi^2_c>_L-<|\phi_c|>^2\right]$ versus the
scaling variable $ t = |z-z_c|L^{1/\nu}$, where $\beta$, $\gamma$ and
$\nu$ are the critical exponents of the order parameter, susceptibility
and correlation length respectively.  On a double logarithmic plot we
obtained data collapses and the 2D--Ising asymptotic large--$t$ behavior
for all cases studied by utilizing the critical exponents of the 2D-Ising
universality class ($\beta=1/8$, $\gamma=7/4$ and $\nu=1$); examples are
shown in Figs.~9, 10, 11. These data, in conjunction with the cumulant
intersection value being independent of $M$ and close to the accepted
value for the 2D Ising class indicate that independent of the number of
components $M$ the transition from the disordered to the crystal phase
belongs to the 2D Ising universality class.

We now discuss the phase transitions to the demixed phase.  The phase
transition was analyzed by a study of the order parameter distribution
$P_L(\phi_d)$, the density $\rho(z)$ and free energy integration.  In
Fig.~12 we show the density of the system versus $z$ for different values
of $M$. The density is an increasing function of $z$ and approaches for
large $z$ the asymptotic form $\rho=z/(1+z)$ which describes the behavior
of the system with $M=1$. In the demixed phase one particle type is
dominant and so the system properties are close to that of a one component
system.  In general the density in the demixed phase exceeds the value
$1/2$, for $M>6$ we observe a direct first order transition from the
crystal to the demixed phase, with a finite jump in the density at the
transition fugacity $z_d$.  For large $M$ the density should jump from
$\rho \approx 1/2$ to $\rho \approx z_d/(1+z_d)$. In the simulations we
find a hysteresis region around $z_d$ going approximately between these
two values when increasing and decreasing the fugacity. In
cases of a small hysteresis region with extent of less than $|z-z_d| <
0.1$ the middle $z$--value of this region was taken as the transition
value $z_d$. In cases of a pronounced hysteresis region we located the
transition fugacity $z_d$ by two independent methods. (i) We integrated
the free energy from $z=0$ to $z_d$ and from $z = \infty$ to $z_d$, where
the limiting free energies are known. For the low fugacity free energy we
obtain:
\begin{equation}
p_l(z) 
%= - \log z 
= - \int_0^z dz_1 \rho(z_1) / z_1
\label{eq.pl}
\end{equation}
For the high fugacity free energy we obtain:
\begin{equation}
p_h(z) = - \int_{z_l}^z dz_1 \rho(z_1) / z_1 - \log(1+z_l)
\label{eq.ph}
\end{equation}
where $z_l$ is chosen that large (typically $z_l=50$) that $\rho(z_l)
\approx z_l/(1+z_l)$. The intersection value $z_d$, where $p_l(z_d)=
p_h(z_d)$ was one estimate for the value of the fugacity where the first
order transition to the demixed phase takes place, see Fig.~13.  (ii)
Another independent estimate for $z_d$ was obtained by the procedure of
searching for relative stability of one of the two phases during a
simulation starting from configurations with both phases present in
parallel slices extending over the length of the simulation box, see
Fig.~13.  Both independent methods gave, within 5\% uncertainty, the same
numerical values for $z_d$.  The resulting phase transition values of
$z_d$ are shown in Fig.~7.  We note that with increasing number of
components, the transition fugacities approach the exact asymptotic line
$M=z+2-1/z$, see section~II.

For the cases $M\geq 5$ we find a jump in the density at $z_d$ so that we
classify these transitions as first order transitions.  For $M \leq 4$ we
do not observe a jump in the density at $z_d$.  In these cases the
transition was located by the cumulant intersection method, see above,
where the order parameter $\phi_d$ is the order parameter of the demixed
phase and $\phi_c$ has to be replaced by $\phi_d$ in Eqs.(\ref{cmoments})
and (\ref{ccumu}).  A finite size scaling analysis shows that the data for
the order parameter and the susceptibility of the phase transitions for
$M\leq 4$ are consistent with the 2D $M$-state Potts universality class,
see Figs.~14, 15. For the susceptibility and $M=4$ we obtain a much better
scaling behavior compared to an analysis assuming the 2D--Ising exponents
to be valid, however we note that the ratios for $\beta/\nu$ and
$\gamma/\nu$ of the Potts classes~\cite{Baxter} agree with the 2D Ising
class values within a few per cents, so that a clear distinction between
these classes based on our numerical data is difficult.  Recently the case
$M=2$ was studied with Monte Carlo methods~\cite{Dick} where evidence for
the 2D--Ising universality class was found, in agreement with our
findings.

%In Fig.~1 we also compare the Monte Carlo results with the results of
%exact calculations on the Bethe lattice with coordination number
%$q=4$. The topology of the phase diagrams agree but the values of the
%transition points are slightly different.

%%%%%%%%%%%%%%%%%%%%%%%
%\section{discussion}

%\section{summary and conclusions}

\acknowledgements We thank L. Chayes and R. Kotecky for useful discussion
about their work~\cite{CKS}. P.N. acknowledges support from the Deutsche
Forschungsgemeinschaft (Heisenberg foundation), the computations were
carried out at the CRAY-YMP of the RHRK Kaiserslautern.  The work of
J.L.L. and A.M. was supported by NSF Grant DMR 92--13424.  The work of
L.\v S. was supported by NSF Grant CHE 92--17893.

%\end{document}

%%%%%%%%%%%%%%%%%%%%%%%%%%%%%%%%%%%%%%%%%%%%%%%%%%%%%%%%%%%%%%%%%%%%%%%%%%%
%%%%%%%                       Appendices
%%%%%%%%%%%%%%%%%%%%%%%%%%%%%%%%%%%%%%%%%%%%%%%%%%%%%%%%%%%%%%%%%%%%%%%%%%%

%%\appendix
%%\section{}
%%\label{app1}

%%%%%%%%%%%%%%%%%%%%%%%%%%%%%%%%%%%%%%%%%%%%%%%%%%%%%%%%%%%%%%%%%%%%%%%%%%%
%%%%%%%%                       FIGURES
%%%%%%%%%%%%%%%%%%%%%%%%%%%%%%%%%%%%%%%%%%%%%%%%%%%%%%%%%%%%%%%%%%%%%%%%%%%

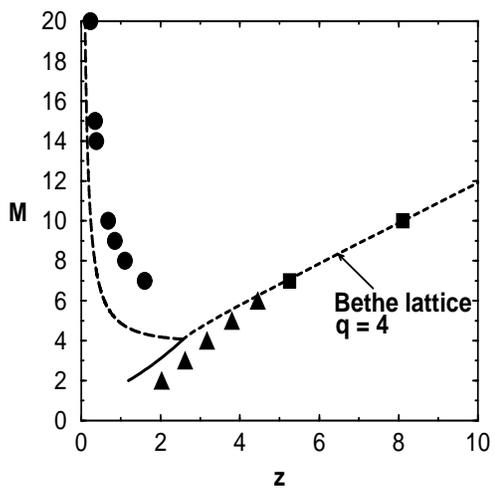
\begin{figure}
%\begin{center}
%\begin{picture}(70,70)
%\put(0,0){\psfig{figure=ffffig14.ps,width=70mm,height=70mm}}
%\end{picture}
\caption[]{
Phase diagram in the $M$--$z$ plane for a square lattice (MC) and for a
Bethe lattice with coordination
number $q=4$ .
Lines:  Exact results for the Bethe lattice for the transition lines from the
gas phase to the crystal phase (dashed line), from the gas to the demixed
phase (full line) and from the crystal to the demixed phase (dotted line).
Symbols for MC: Transition points from the
gas phase to the crystal phase (circles), from the gas to the demixed
phase (triangles) and from the crystal to the demixed phase (squares).}
%\end{center}
\label{FIG14}
\end{figure}

\unitlength1mm
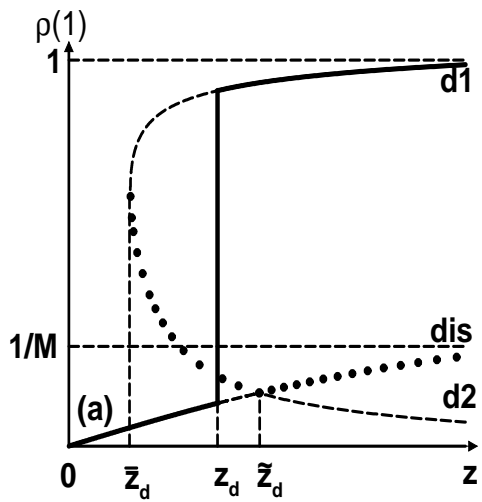
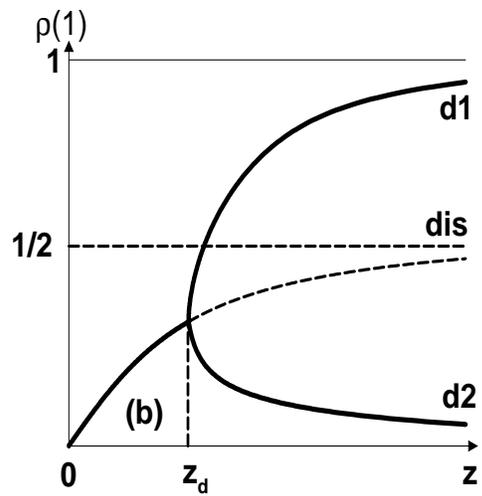
\begin{figure}[hbt]
%\begin{center}
%\begin{picture}(70,70)
%\put(0,0){\psfig{figure=ffffig2a.ps,width=70mm,height=70mm}}
%\end{picture}
%\begin{picture}(70,70)
%\put(0,0){\psfig{figure=ffffig2b.ps,width=70mm,height=70mm}}
%\end{picture}
\caption[]{A schematic plot of the interplay among the disordered (dis),
demixed 1 (d1) and demixed 2 (d2) Bethe solutions in the 
$[z,\rho(1)]$ plane for a) $M>2$, b) $M=2$; the solution with the
lowest, intermediate and highest free energy is depicted by
the solid, dashed and dotted line, respectively.}
%\end{center}
\label{FIG2}
\end{figure}

\unitlength1mm
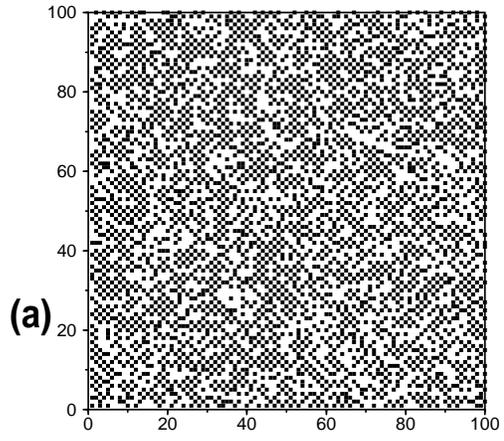
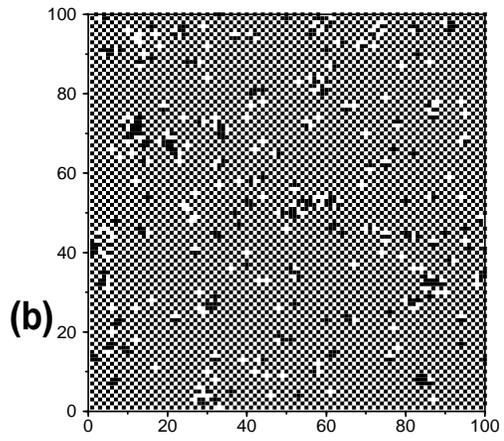
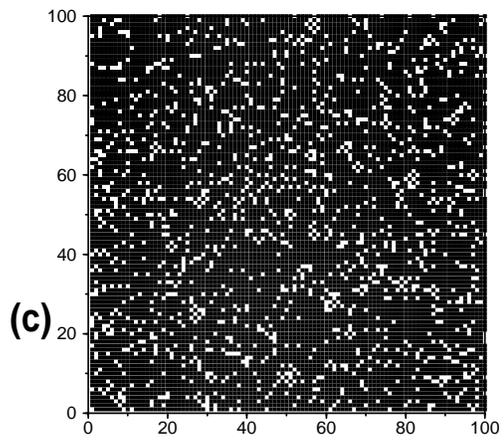
\begin{figure}
%\begin{center}
%\begin{picture}(70,70)
%\put(0,0){\psfig{figure=f1a.ps,width=70mm,height=70mm}}
%\put(0,0){\psfig{figure=fa.ps,width=70mm,height=70mm}}
%\end{picture}
%\begin{picture}(70,70)
%\put(0,0){\psfig{figure=f1b.ps,width=70mm,height=70mm}}
%\put(0,0){\psfig{figure=fb.ps,width=70mm,height=70mm}}
%\end{picture}
%\begin{picture}(70,70)
%\put(0,0){\psfig{figure=f1c.ps,width=70mm,height=70mm}}
%\put(0,0){\psfig{figure=fc.ps,width=70mm,height=70mm}}
%\end{picture}
\caption[]{
Typical configurations for $M=9$ and  (a) $z=0.1$, (b) $z=5$ and (c) $z=8.5$.
}
%\end{center}
\label{FIG1}
\end{figure}

%\twocolumn

\unitlength1mm
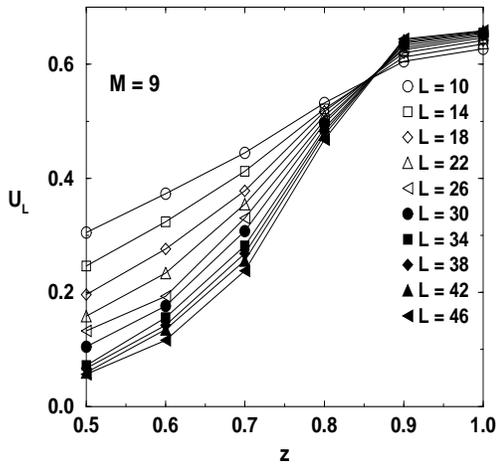
\begin{figure}
%\begin{center}
%\begin{picture}(70,70)
%\put(0,0){\psfig{figure=ffig2.ps,width=70mm,height=70mm}}
%\end{picture}
\caption[]{
Cumulant $U_L$ versus $z$ for $M=9$. The different symbols refer to
different subsystem sizes as indicated in the figure.
Lines are for visual help.
}
%\end{center}
\label{FIG2b}
\end{figure}

%\unitlength1mm
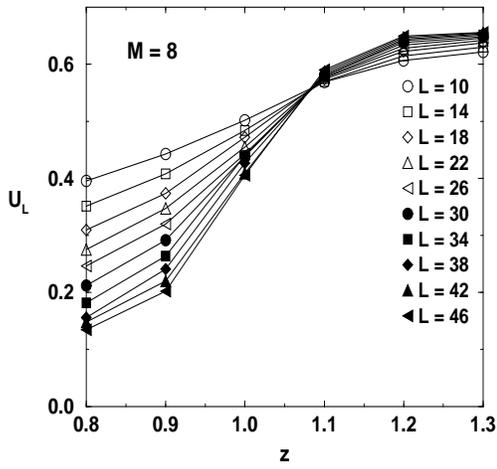
\begin{figure}
%\begin{center}
%\begin{picture}(70,70)
%\put(0,0){\psfig{figure=ffig3.ps,width=70mm,height=70mm}}
%\end{picture}
\caption[]{
Cumulant $U_L$ versus $z$ for $M=8$. The different symbols refer to
different subsystem sizes as indicated in the figure.
Lines are for visual help.
}
%\end{center}
\label{FIG3}
\end{figure}

%\unitlength1mm
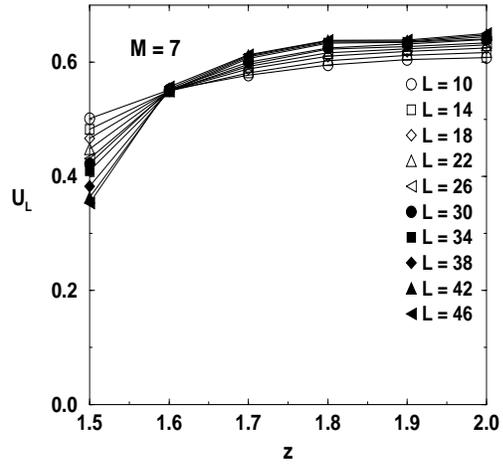
\begin{figure}
%\begin{center}
%\begin{picture}(70,70)
%\put(0,0){\psfig{figure=ffig4.ps,width=70mm,height=70mm}}
%\end{picture}
\caption[]{
Cumulant $U_L$ versus $z$ for $M=7$. The different symbols refer to
different subsystem sizes as indicated in the figure.
Lines are for visual help.
}
%\end{center}
\label{FIG4}
\end{figure}

%\newpage
%\onecolumn
\unitlength1mm
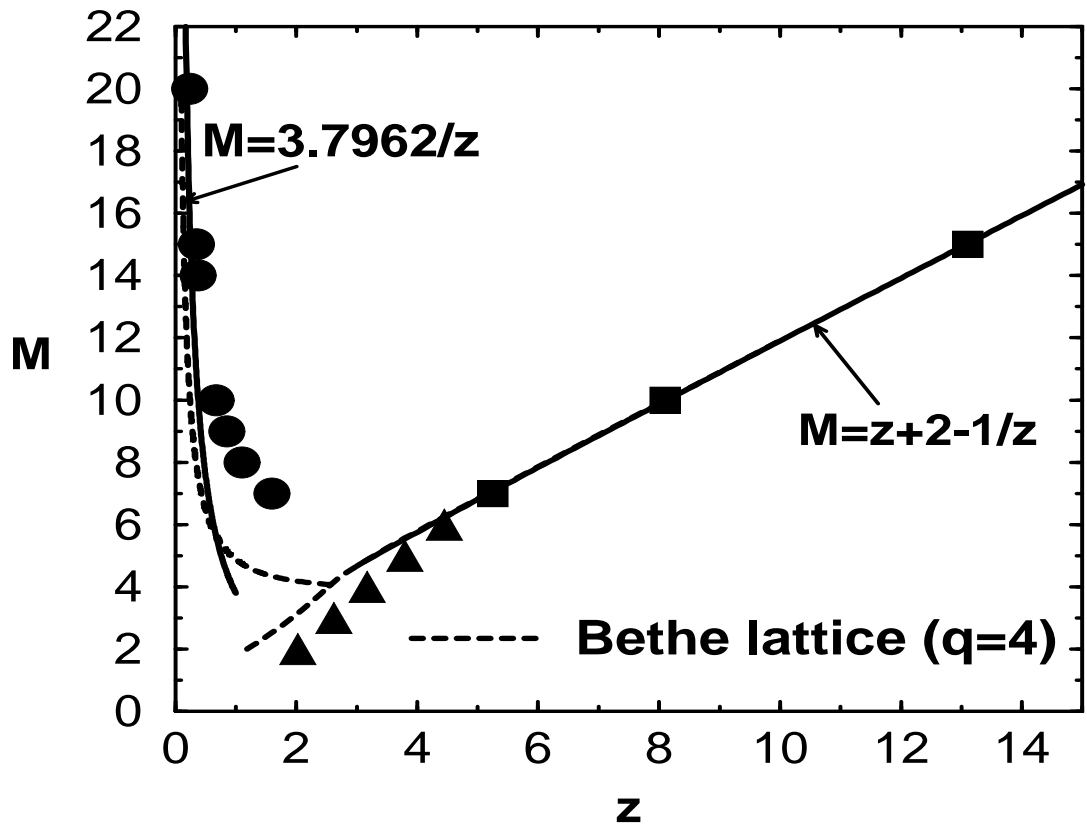
\begin{figure}[hbt]
%\begin{center}
%\begin{picture}(160,120)
%\put(0,0){\psfig{figure=ffffig5.ps,width=160mm,height=120mm}}
%\end{picture}
\caption[]{
Phase diagram in the $M$--$z$ plane for a square lattice (MC).
Full lines:  Asymptotic lines for the phase transitions in the high fugacity 
region, $M=z+2-1/z$, and for the transition in the low fugacity region, 
$Mz=3.7962$.
Dashed lines: Results from the computation on the Bethe lattice with
coordination number $q = 4$.
Symbols for MC: Transition points from the
gas phase to the crystal phase (circles), from the gas to the demixed
phase (triangles) and from the crystal to the demixed phase (squares).
}
%\end{center}
\label{FIG5}
\end{figure}

%\twocolumn
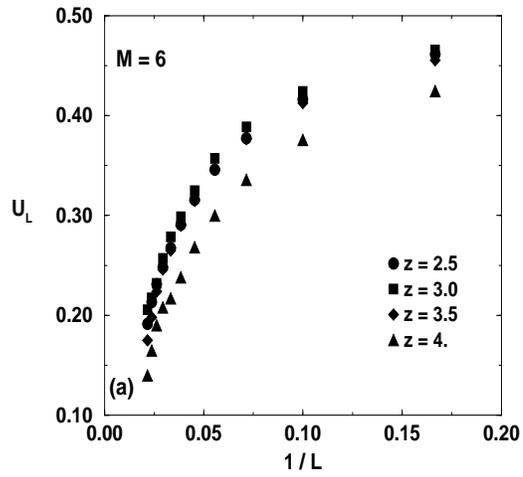
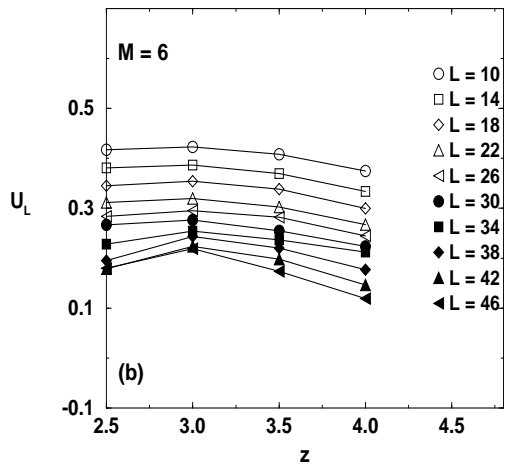
\begin{figure}
%\begin{center}
%\begin{picture}(70,70)
%\put(0,0){\psfig{figure=ffig6a.ps,width=70mm,height=70mm}}
%\end{picture}
%\begin{picture}(70,70)
%\put(0,0){\psfig{figure=ffig6b.ps,width=70mm,height=70mm}}
%\end{picture}
\caption[]{
(a) Cumulant $U_L$ versus inverse system size for $M = 6$ and
various values of $z$.
(b) Cumulant $U_L$ versus $z$ for $M=6$. The different symbols refer to
different subsystem sizes as indicated in the figure.
Lines are for visual help.
}
%\end{center}
\label{FIG6}
\end{figure}

%%%%%%%%%%%%%%%%%%%%%%%
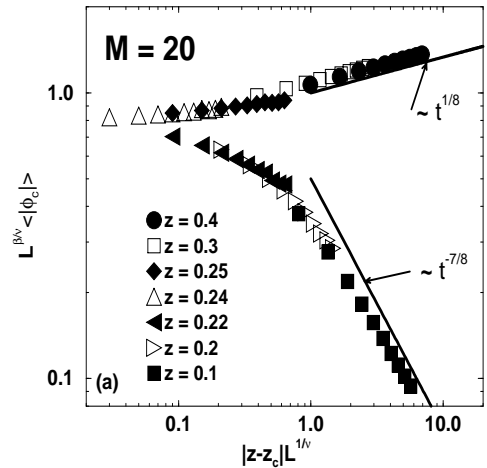
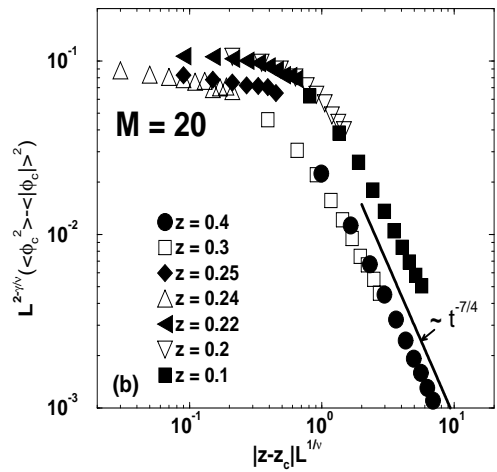
\begin{figure}
%\begin{center}
%\begin{picture}(70,70)
%\put(0,0){\psfig{figure=ffig7a.ps,width=70mm,height=70mm}}
%\end{picture}
%\begin{picture}(70,70)
%\put(0,0){\psfig{figure=ffig7b.ps,width=70mm,height=70mm}}
%\end{picture}
\caption[]{
Scaling functions of the order parameter (a) and the order parameter 
susceptibility (b) for $M=20$ 
utilizing the 2D--Ising critical exponents.
Lines indicate the asymptotic power law behavior 
with the 2D--Ising critical exponents
($t=|z-z_c|L^{1/\nu}$).
}
%\end{center}
\label{FIG7}
\end{figure}

%%%%%%%%%%%%%%%%%%%%%%%

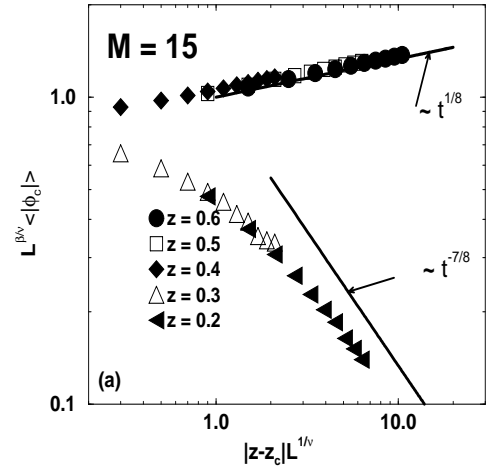
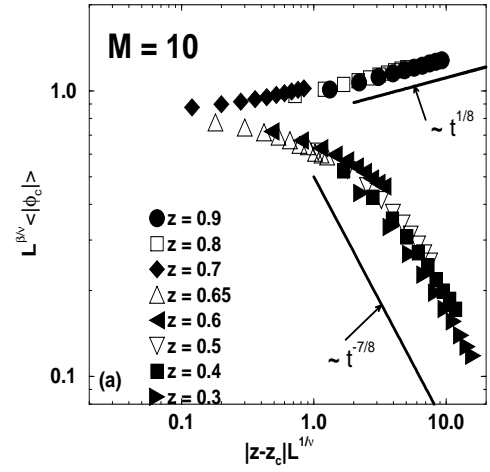
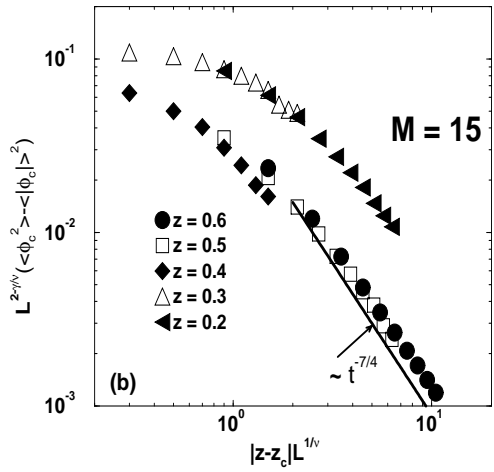
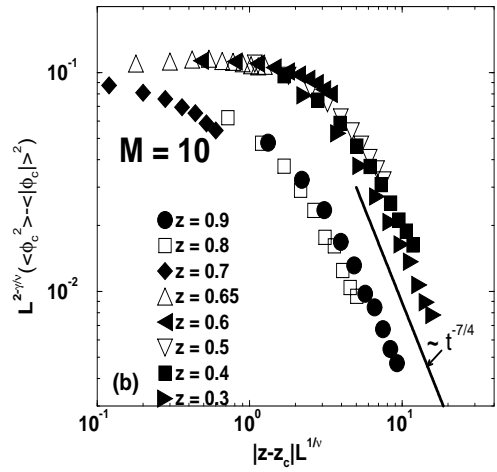
\begin{figure}
%\begin{center}
%\begin{picture}(70,70)
%\put(0,0){\psfig{figure=ffig8a.ps,width=70mm,height=70mm}}
%\end{picture}
%\begin{picture}(70,70)
%\put(0,0){\psfig{figure=ffig8b.ps,width=70mm,height=70mm}}
%\end{picture}
\caption[]{
Scaling functions of the order parameter (a) and the order parameter 
susceptibility (b) for $M=15$
utilizing the 2D--Ising critical exponents.
Lines indicate the asymptotic power law behavior 
with the 2D--Ising critical exponents
($t=|z-z_c|L^{1/\nu}$).
}
%\end{center}
\label{FIG8}
\end{figure}

%%%%%%%%%%%%%%%%%%%%%%%

\begin{figure}
%\begin{center}
%\begin{picture}(70,70)
%\put(0,0){\psfig{figure=ffig9a.ps,width=70mm,height=70mm}}
%\end{picture}
%\begin{picture}(70,70)
%\put(0,0){\psfig{figure=ffig9b.ps,width=70mm,height=70mm}}
%\end{picture}
\caption[]{
Scaling functions of the order parameter (a) and the order parameter 
susceptibility (b) for $M=10$
utilizing the 2D--Ising critical exponents.
Lines indicate the asymptotic power law behavior 
with the 2D--Ising critical exponents
($t=|z-z_c|L^{1/\nu}$).
}
%\end{center}
\label{FIG9}
\end{figure}

%%%%%%%%%%%%%%%%%%%%%%%

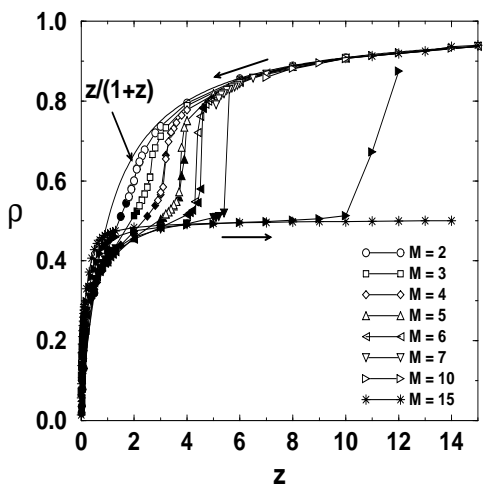
\begin{figure}
%\begin{center}
%\begin{picture}(70,70)
%\put(0,0){\psfig{figure=ffig10.ps,width=70mm,height=70mm}}
%\end{picture}
\caption[]{
Density of the system versus $z$ for different number of components $M$,
symbols refer to Monte Carlo results, the connecting lines are for visual help.
The full (open) symbols refer to the MC results obtained by
starting with configurations previosly obtained for lower (higher) fugacities. 
The full line equals $\rho=z/(1+z)$.
}
%\end{center}
\label{FIG10}
\end{figure}

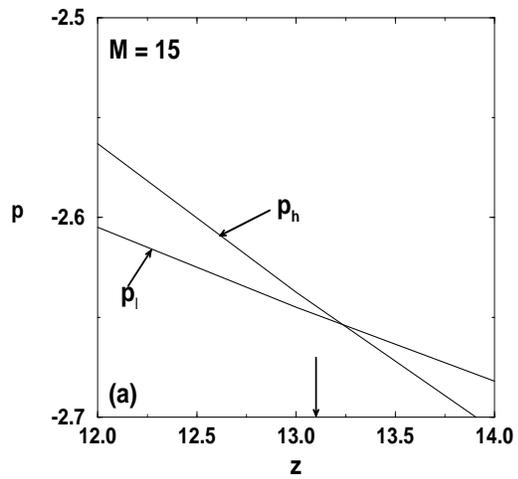
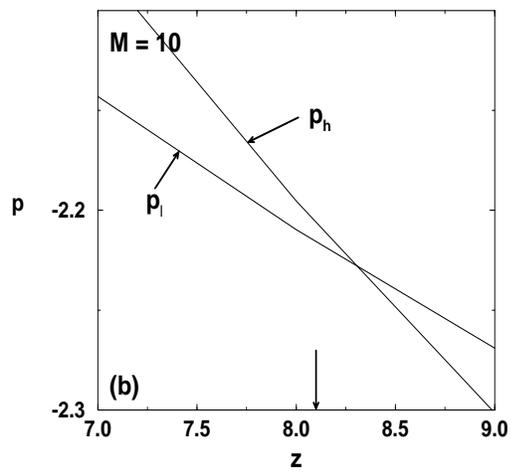
\begin{figure}
%\begin{center}
%\begin{picture}(70,70)
%\put(0,0){\psfig{figure=ffig11a.ps,width=70mm,height=70mm}}
%\end{picture}
%\begin{picture}(70,70)
%\put(0,0){\psfig{figure=ffig11b.ps,width=70mm,height=70mm}}
%\end{picture}
\caption[]{
Free energy $p_l(z)$ and $p_h(z)$ from thermodynamic integration
for (a) $M = 15$ and (b) $M = 10$. The arrow at the $z$--axes indicates
the transition value found by the phase stability study, see text.
}
%\end{center}
\label{FIG11}
\end{figure}

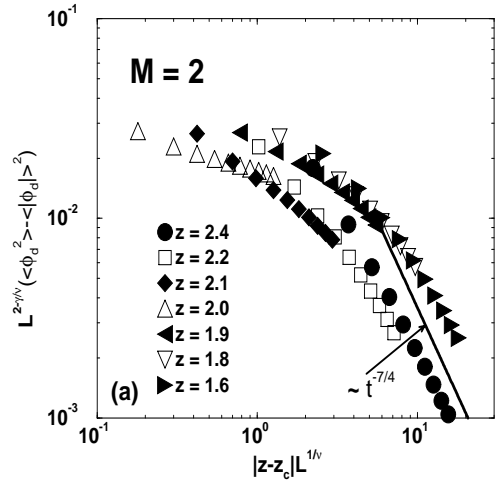
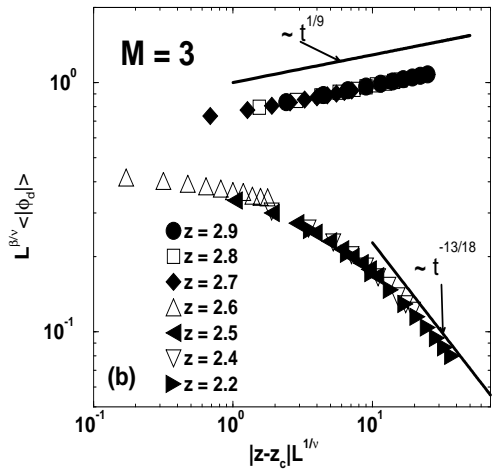
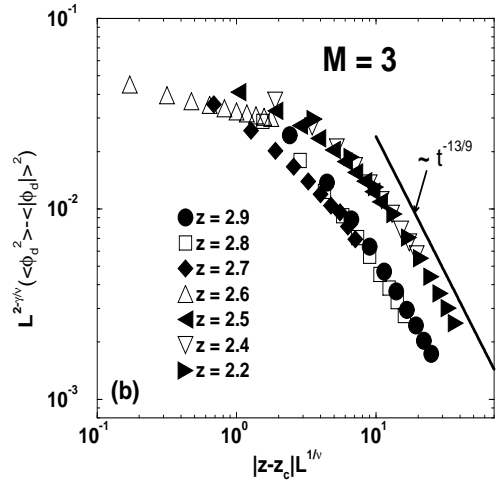
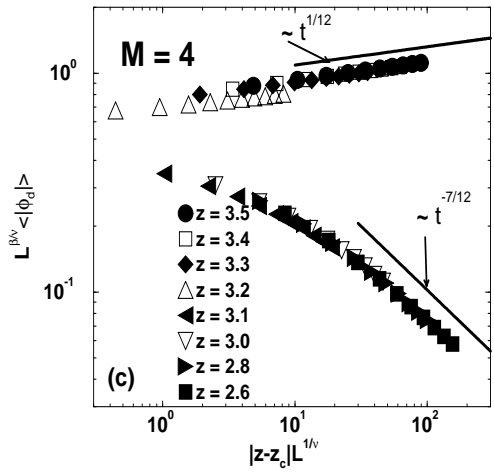
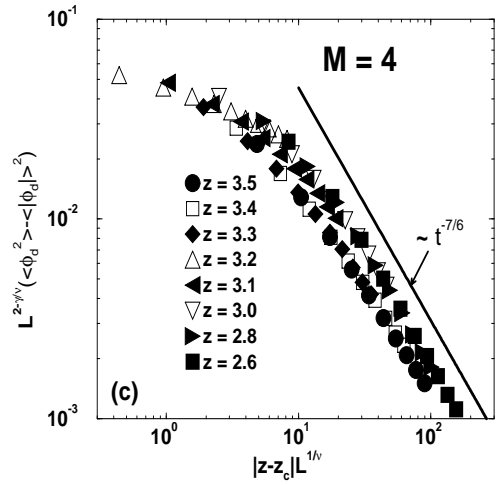
\begin{figure}
%\begin{center}
%\begin{picture}(70,70)
%\put(0,0){\psfig{figure=ffig12a.ps,width=70mm,height=70mm}}
%\end{picture}
%\begin{picture}(70,70)
%\put(0,0){\psfig{figure=fffig12b.ps,width=70mm,height=70mm}}
%\end{picture}
%\begin{picture}(70,70)
%\put(0,0){\psfig{figure=ffig12c.ps,width=70mm,height=70mm}}
%\end{picture}
\caption[]{
Scaling function of the order parameter for $M=2$ (a), $M=3$(b) and $M=4$ (c)
using the 2D $M$-state Potts critical exponents. 
Lines indicate the asymptotic power law behavior 
with the 2D--$M$-state Potts critical exponents
($t=|z-z_c|L^{1/\nu}$).
}
%\end{center}
\label{FIG12}
\end{figure}
\begin{figure}
%\begin{center}
%\begin{picture}(70,70)
%\put(0,0){\psfig{figure=ffig13a.ps,width=70mm,height=70mm}}
%\end{picture}
%\begin{picture}(70,70)
%\put(0,0){\psfig{figure=fffig13b.ps,width=70mm,height=70mm}}
%\end{picture}
%\begin{picture}(70,70)
%\put(0,0){\psfig{figure=ffig13c.ps,width=70mm,height=70mm}}
%\end{picture}
\caption[]{
Scaling function of the order parameter susceptibility
 for $M=2$ (a), $M=3$(b) and $M=4$ (c)
using the 2D $M$-state Potts critical exponents. 
Lines indicate the asymptotic power law behavior 
with the 2D--$M$-state Potts critical exponents
($t=|z-z_c|L^{1/\nu}$).
}
%\end{center}
\label{FIG13}
\end{figure}

%%%%%%%%%%%%%%%%%%%%%%%%%%%%%%%%%%%%%%%%%%%%%%%%%%%%%%%%%%%%%%%%%%%%%%%%%%%
%%%%%%%                       TABLES
%%%%%%%%%%%%%%%%%%%%%%%%%%%%%%%%%%%%%%%%%%%%%%%%%%%%%%%%%%%%%%%%%%%%%%%%%%%

% \begin{table}
% \caption{}
% \label{}
% \end{tabular}
% \end{table}
%
%\end{multicols}
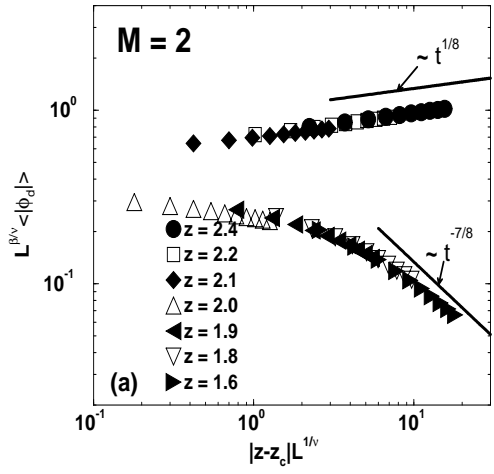
\end{document}